\documentclass[preprint,nofootinbib,aps,preprintnumbers,amsmath,amssymb]{revtex4}
\usepackage{graphicx}
\usepackage{dcolumn}
\usepackage{bm}
\usepackage{natbib}

\newcommand{\be}{\begin{equation}}
\newcommand{\ee}{\end{equation}}
\newcommand{\beq}{\begin{equation}}
\newcommand{\eq}{\end{equation}}
\newcommand{\bea}{\begin{eqnarray}\displaystyle}
\newcommand{\eea}{\end{eqnarray}}
\newcommand{\ea}{\end{eqnarray}}

\newcommand{\p}{\partial}
\newcommand{\s}{\sigma}

\newcommand{\vol}{V_0^{(7)}}

\begin{document}

\title{Dilaton Dynamics from Production of Tensionless Membranes}

\author{Sera Cremonini}
\email{sera@het.brown.edu}
\affiliation{Physics Department, Brown University\\
Providence, RI 02906}
\author{Scott Watson}
\email{watsongs@physics.utoronto.ca}
\affiliation{Physics Department, University of Toronto \\
Toronto, ON}

\date{\today}

\begin{abstract}
In this paper we consider classical and quantum corrections to
cosmological solutions of $11$D SUGRA coming from dynamics of
membrane states. We first consider the supermembrane spectrum
following the approach of Russo and Tseytlin for consistent
quantization. We calculate the production rate of BPS membrane
bound states in a cosmological background and find that such
effects are generically suppressed by the Planck scale, as
expected. However, for a modified brane spectrum possessing
enhanced symmetry, production can be finite and significant. We
stress that this effect could not be anticipated given only a
knowledge of the low-energy effective theory. Once on-shell,
inclusion of these states leads to an attractive force pulling the
dilaton towards a fixed point of S-duality, namely $g_s=1$.
Although the SUGRA description breaks down in this regime,
inclusion of the enhanced states suggests that the center of
M-theory moduli space is a dynamical attractor. Morever, our
results seem to suggest that string dynamics does indeed favor a
vacuum near fixed points of duality.
\end{abstract}

\maketitle
\tableofcontents
\newpage

\section{Introduction}
Low-energy descriptions of string theory generically predict the
existence of a large number of scalar fields,
or moduli, which are associated with the size and shape of the extra
dimensions,
as well as the position and orientation of any branes present in the
theory.
These moduli are of interest for a number of reasons.
From a theoretical point of view, the different vacuum expectation
values (VEVs) of these
fields correspond to different choices for the string vacuum, leading
to an indeterminacy of the theory.
From a more phenomenological viewpoint, light scalars can have a profound
effect
on both the early and late universe. If not fixed, these moduli can
lead to a period of
inflation, modifications of fundamental constants, and violations of
the equivalence principle.
If fixed, one must worry about the mass scales involved and the
effects of the resulting relic density on cosmological
observations, e.g. on the cosmic microwave background
or Big Bang nucleosynthesis.

Recently it has been argued \cite{Vafa:2005ui} (see also
\cite{Arkani-Hamed:2006dz}) that, by taking low-energy
supergravity (SUGRA) as the effective description for string theory, we
may
miss certain crucial aspects of the underlying theory.
As an example, when attempting to understand the vacuum
structure of the theory one
should remember that moduli space must be of finite size in order to
have a realistic theory of gravity (finite $G_N$).
Other crucial aspects that we wish to address in this paper are the
role of dualities and the importance of dynamics.
That is, as background fields evolve the effective
mass of heavy states that are outside the realm of the low-energy
theory can change.
In fact, near points of enhanced symmetry (ESPs), frequently associated
with dualities, these additional states can become massless and
therefore play a vital role in the low-energy theory.
It is important that these states lie beyond the naive low-energy
SUGRA description and, without an underlying knowledge of the
fundamental theory,
effects associated with these additional degrees of freedom (such as
particle creation and radiative corrections) would be missed.
Indeed, recent work suggests that including these effects can have
important and interesting effects both in string theory and cosmology
\cite{Silverstein:2003hf,Kofman:2004yc,Watson:2004aq,Alishahiha:2004eh}.

In this paper we will consider cosmological solutions of $11$D SUGRA
and ask what corrections, if any, result from the dynamics of the
moduli (i.e., dilaton and radii).  Focusing on the case of particle
production, we show that production of BPS membrane bound states,
whose mass depends on the evolution of the moduli, is generally
suppressed by the Planck scale.  This is expected and assures us of
the validity of the low-energy effective theory.  However, in the
special case of membranes that become tensionless near fixed points of
duality, we find that significant production can occur.  We are unaware of an explicit construction of such states, however we anticipate their existence given ubiquitous examples in the
lower-dimensional string theory case.

We find that by including enhanced states in the low-energy theory, the
center of the M-theory moduli space becomes an attractor, and the
evolution of the moduli tends toward fixed points of duality.  In the
case of the dilaton, which can be taken to be $R_{11}$ in the $11$D
theory, we find that the evolution leads to the region near $g_s=1$,
i.e. strong coupling.
Naively trusting the
equations of motion in this region, we find that all moduli will in
fact become trapped and we recover a three dimensional radiation
dominated universe at late times. Although we expect additional
corrections to arise near the regime of strong coupling, it is
intriguing that our results seem to provide an
explicit example of the belief (\cite{Brustein:2002xf} and
\cite{Dine:1998qr}) that a string vacuum consistent with gravity and
particle phenomenology should lie near fixed points of duailty.

The paper is organized as follows.  In Section {\ref{MthStates} we
review a scheme developed by Russo and Tseytlin for quantizing the
supermembrane in special limits.  This will allow us to obtain the BPS
spectrum of membrane bound states, which we will utilize throughout
the paper.
In Section \ref{ClTrapping} we briefly discuss cosmology with these
states and show that they can lead to stabilization, but we also point out
that
such considerations seem inconsistent, since they lie outside the scope
of the effective theory.  In Section \ref{QmTrapping} we consider the
possibility of producing membrane bound states within a cosmological
background of $11$D SUGRA.  After reviewing the basic formalism for
calculating production, we find an expression for the production rate
and find that membrane bound states suffer exponential suppression, as
expected.  However, we also find that, for a special class of states
possessing enhanced symmetry, significant production can result.
In Section \ref{cosmo} we consider the modified effective theory in
the presence of the produced states, and determine the new evolution
of the moduli.  We find an attractor behavior towards the center of
moduli space, and trusting the equations in the center region leads to
trapping of the moduli.  We conclude with a discussion of the limitations of our approach and future work in progress.

\section{M-theory and the Supermembrane \label{MthStates}}

In this section we review progress that has been made in understanding M-theory
as a fundamental theory of membranes.
The goal will be to gain an understanding of the spectrum of the fundamental theory
beyond the low-energy effective description of $11$D SUGRA.
We will be concerned with finding membrane states whose mass depends explicitly
on the low-energy moduli, namely the dilaton and radii of the extra dimensions.

We begin by reviewing the work of \cite{Russo:1996if}, where
the authors were able to clarify the $11$D origin of many
lower-dimensional solutions, as well as identify states of the supermembrane
with known string theory configurations.
Much effort has been devoted to understanding composite BPS
configurations of branes in $11$D.
We will concentrate in particular on non-threshold (non-zero
binding energy) bound states, the canonical example of which (in string theory) is
the $(p,q)$ string, a bound
state of a NS-NS string and a R-R string in type IIB theory.

In \cite{Schwarz:1995du,Schwarz:1996bh}, it was
suggested that the $(p,q)$ string states should be related to the
BPS states of a wrapped M$2$-brane in $11$D. This was supported by a
comparison of the {\it zero-mode} contributions of the membrane to the
string mass spectra, where agreement was found. However, matching at
the oscillator level was not shown, due to the intrinsic
difficulties associated with the quantization of the supermembrane.
This difficulty was overcome and
the {\it oscillating} contributions to the spectrum were
obtained in \cite{Russo:1996if}, where supermembrane
quantization was achieved in a specific limit in which the theory
dramatically simplifies.

Some of the M-theory
solutions studied in \cite{Russo:1996if} can have interesting
cosmological consequences, as will become apparent in the next section.
Specifically, we are interested in certain M-theory states
which have a classical supergravity description in terms of
a bound state of an M$2$-brane wrapping a $T^2$
and of a gravitational wave propagating along one of the torus directions.
We will start by presenting the 11D SUGRA solution for such configurations, and show
that in a certain limit it can be reduced to the type IIB $(p,q)$ string.
Since we are primarily interested in obtaining the mass spectrum of these states, we will
then present their interpretation in supermembrane theory.
Specifically, we will outline the derivation of the spectrum of such
solutions as given in \cite{Russo:1996if}, where it was first confirmed that
the mass of an oscillating membrane does indeed agree with that of the $(p,q)$ string.
Finally, at the end of this section we attempt to motivate a new class of states possessing enhanced symmetry.  We will save an explicit construction of such states for future work, here only comparing their spectrum to the know BPS states of the supermembrane.  We will see in future sections that such states are the only possibility for membrane production in a low-energy theory.

\subsection{D=11 SUGRA Solutions}

We present a concise introduction to the 11D origin of certain
non-marginal BPS configurations of type II string theories.
We consider the 11-dimensional space $M^9 \times T^2$.
The torus coordinates are labeled by $(y_1,y_2)$ and have periods
$(2\pi R_9,2\pi R_{11})$, while the spatial coordinates of $M^9$ are denoted by $x_i$, $i=1,\ldots,8$.
We are ultimately interested in considering a bound state of an M$2$ brane,
which wraps the $T^2$, and of a gravitational wave propagating in
the $y_2$ direction. We will briefly show how to construct such states
from simpler solutions.

The 11D solution representing a gravitational wave moving along $y_1$ is given by
\bea
ds^2 &=& -dt^2 + dy_1^2 + dy_2^2 + W (d t+dy_1)^2+ dx_i dx_i \, ,\nonumber \\
C_{\mu \nu \rho} &=& 0, \; \; \; W=\frac{\tilde{Q}}{r^6}, \; \; \;
r^2=x_i x_i.
\ea
Since $y_1$ is periodic, the charge $\tilde{Q}$ must be quantized
in units of $1/R_9$; with the correct normalization factor, the
charge becomes $\tilde{Q}=c_0 \frac{l}{R_9}$.
This solution is found to preserve $1/2$ of the supersymmetries.
The other basic 11D object that we will need is the 2-brane, with
solution \cite{Russo:1996if},
\bea
ds^2 &=& H_2^{-\frac{2}{3}} \, \left[-dt^2 + dy_1^2 + dy_2^2 \right] + H_2^{\frac{1}{3}} dx_i
dx_i \, , \nonumber \\
C_3 &=& H_2^{-1} dt \wedge dy_1 \wedge dy_2, \;\;\;
H_2=1+\frac{Q}{r^6}.
\ea
The charge in this case is given by $Q=c_0 \frac{w_0 R_9}{\alpha'}$, where
$w_0$ denotes winding around the target torus (we will define
$w_0$ more precisely in the next section).

Combining the 2-brane solution with a gravitational wave moving in
an {\it arbitrary} direction gives
\bea
ds^2 &=& H_2^{-\frac{2}{3}} \, \bigl[-dt^2 + dy_1^2 + dy_2^2 + W (d t-dz_1)^2\bigr]+ H_2^{\frac{1}{3}} dx_i
dx_i\, , \nonumber \\
C_3 &=& H_2^{-1} dt \wedge dy_1 \wedge dy_2, \; \; \; z_1=y_1
\cos{\theta}+y_2\sin{\theta}\, ,
\ea
where the gravitational wave charge is modified
in the following way:
\bea
W&=&\frac{\tilde{Q}_q}{r^6}, \; \; \; \tilde{Q}_q=c_0\sqrt{\frac{l_9^2}{R_9^2}+\frac{l_{11}^2}{R_{11}^2}}
=\tilde{Q}\sqrt{q_1^2+q_2^2 \tau_2^2},\\
\text{with}&& \; \; \; \cos{\theta}=\frac{q_1}{\sqrt{q_1^2+q_2^2 \tau_2^2}}, \; \; \; \; \;
\tau_2=\frac{R_9}{R_{11}}.
\ea
After appropriate dimensional reductions and applications of dualities,
this solution can be identified with a boosted BPS bound state of
an $F1$ string and $D0$ brane in Type IIA string theory, and with
a boosted $(p,q)$ string in Type IIB.

We are specifically interested in the 2-brane + wave solution with
a boost along $y_2$ only (with $\cos{\theta}=0$  and $\sin{\theta}=1$),
giving \cite{Russo:1996if},
\bea
ds^2 &=& H_2^{-\frac{2}{3}} \, \bigl[-dt^2 + dy_1^2 + dy_2^2 + W (d t-dy_2)^2\bigr]+ H_2^{\frac{1}{3}} dx_i
dx_i, \nonumber \\
C_3 &=& H_2^{-1} dt \wedge dy_1 \wedge dy_2,
\ea
where
\bea
H_2&=&1+\frac{Q}{r^6}, \; \; \; r^2=x_i x_i,  \; \; \; Q= (4\pi^2 T_3 c_0 w_0 R_9)\, R_{11}, \\
W&=& \frac{\tilde{Q}}{r^6},  \; \; \; \tilde{Q}=c_0
\frac{l_{11}}{R_{11}},
\ea
and we used the definition of $\alpha'$. This state is BPS, preserving $1/4$ of the supersymmetries.

To obtain the mass spectrum of the M-theory membrane that
corresponds to this classical supergravity solution, we will make use of
supermembrane theory.
Before proceeding, we would like to stress the advantage of studying
BPS solutions. The BPS condition guarantees that the above classical
SUGRA solutions will exhibit some of the features of the full
quantum (string) theory, since these states are protected from quantum corrections.
That is, if we construct these states in a limit of the theory that is well understood, we can then
extrapolate them to regimes that are less understood,
giving us a partial knowledge of the spectrum of the theory.

\subsection{Supermembrane Mass Spectrum}

The original studies of the physical spectrum of wrapped membranes of toroidal topology
have been performed in \cite{Bergshoeff:1987qx,Duff:1987cs}.
Quantization of the supermembrane is highly non-trivial, and one
might even wonder whether a consistent quantum theory of the
supermembrane can be defined. Addressing the {\it oscillating}
membrane is particularly difficult, since it involves dealing with
a highly non-linear interacting theory.
In the work of \cite{Russo:1996if}, however, quantization was
achieved in an appropriate limit in which the interacting terms
dropped out, and the theory could be solved.
Next, we will outline the arguments of
\cite{Russo:1996if} leading to the mass spectrum of the oscillating
membrane.

We consider a membrane on $\mathcal{M}^9 \times T^2$. The compact
directions are labelled by $X^9$ and $X^{11}$, with radii $R_9,
R_{11}$.
Wrapping the membrane around the toroidal directions
gives
\beq
\label{X1X2}
X^9(\sigma,\rho)= w_9 \, R_9\, \sigma +
\tilde{X}^9(\sigma,\rho)\, , \;\;\;\;\;\;\;\;\;
X^{11}(\sigma,\rho)= w_{11} \, R_{11}\, \rho +
\tilde{X}^{11}(\sigma,\rho)\, ,
\eq
where the (single-valued) functions
$\tilde{X}^9(\sigma,\rho)$ and $\tilde{X}^{11}(\sigma,\rho)$
can be expanded in a complete basis of functions on the torus:
\beq
\label{X1X2tilde}
\tilde{X}^9(\sigma,\rho)=\sqrt{\alpha'}\,\sum_{k,m}X^9_{(k,m)}\,
e^{ik\sigma+im\rho} , \;\;\;\;\;\;
\tilde{X}^{11}(\sigma,\rho)=\sqrt{\alpha'}\,\sum_{k,m}X^{11}_{(k,m)}\,
e^{ik\sigma+im\rho}\, . \eq The constants $\alpha'$ and $T_3$ are
defined as \beq \alpha'=\frac{1}{4 \pi^2 R_{11} T_3}, \; \; \; \;
\; T_3=\frac{l_P^{-3}}{4\pi^2}.
\eq
It will turn out to be convenient to define a
winding number $w_0$ in the following way,
\beq
w_0 =
\frac{1}{(2\pi R_9) (2\pi R_{11})} \int d\sigma d\rho \,
\{X^9,X^{11}\}=w_9 w_{11}, \;\;\;\;\; \{X,Y\}\equiv\p_\sigma X\p_\rho
Y-\p_\rho X\p_\sigma Y, \nonumber
\eq
counting how many times the
membrane is wound around the target torus. A membrane with $w_0
\neq 0$ is stable for topological reasons \cite{Russo:1996ph}, as
will become explicit in the Hamiltonian description. This is a
motivation for wrapping the membrane on $\mathcal{M}^9 \times T^2$,
and not on $\mathcal{M}^{10} \times S^1$, which would give $w_0=0$.

The transverse (single-valued) coordinates $X^i$, $i=1, \ldots ,8$, and their
corresponding canonical momenta can also be expanded on the torus:
\beq
\label{Xi}
X^i(\sigma,\rho)=\sqrt{\alpha'}\,\sum_{k,m}X^i_{(k,m)}\,
e^{ik\sigma+im\rho}, \; \; \; \; \; \; \; \;
P^i(\sigma,\rho)=\frac{1}{4 \pi^2
\sqrt{\alpha'}}\,\sum_{k,m}P^i_{(k,m)}\,
e^{ik\sigma+im\rho}.
\eq
The (bosonic) light-cone Hamiltonian for the
supermembrane is given by
\beq
H_B=2\pi^2 \int d\s d\rho \bigl[(P^{\,c})^{2} + \frac{1}{2}T_3^2 (\{X^c,X^d\})^2
\bigr],
\eq
where $c,d=1,\ldots,10$.
Here we neglect the fermionic sector, which can however be
incorporated.
Making use of the expansions (\ref{X1X2}),(\ref{X1X2tilde}),(\ref{Xi}), and
separating the contributions from the winding modes, one finds
\footnote{For a rigorous and detailed derivation of the Hamiltonian see
\cite{Russo:1996if}. Here we are interested in outlining only the
major steps needed to obtain the mass spectrum.}
\bea
H_B&=&H_0+H_{int}, \nonumber \\
\label{H0}
\alpha' H_0 &=& \frac{R_9^2 w_0^2}{2\alpha'}
+\frac{1}{2}\sum_{k,m} [P_{(k,m)}^a P_{(-k,-m)}^a + \omega^2_{k\,m}X_{(k,m)}^a X_{(-k,-m)}^a]\\
\alpha' H_{int}&=& \frac{1}{4 g^2} \sum_{m,n,p} \sum_{k,l,l'}
(pk'-m'k)(nl-ml')
X_{(l',n)}^a X_{(k,p)}^a X_{(k',m')}^b X_{(l,m)}^b + \nonumber \\
&& + \frac{iw_{11}}{g} \sum_{k,m,n} m k^2 X_{2(0,m)}
X_{(k,n)}^iX_{(-k,-n-m)}^i,
\ea
where
\bea
a,b &=&1, \ldots, 8,11, \;\;\;\; m'=-m-n-p, \; \; \; \; k'=-k-l-l' \nonumber \\
g^2 &\equiv & \frac{R_{11}^2}{\alpha'}=4\pi^2 R_{11}^3 T_3, \; \; \; \; \;
\omega_{k m} =\sqrt{w_{11}^2 k^2+w_9^2m^2\tau_2^2}, \; \; \;
\tau_2=\frac{R_9}{R_{11}}.
\ea
This is clearly a highly non-linear interacting theory.
However, notice that the interacting terms are of order ${\cal O}(\frac{1}{g})$
and  ${\cal O}(\frac{1}{g^2})$. In the large $g$ limit, such terms
are negligible, and can be dropped.
Thus, in the limit $g\rightarrow\infty$, given by
\beq
\label{limit}
R_9,R_{11} \rightarrow \infty, \; \; T_3
\rightarrow 0 \; \; \; \; (\text{holding} \;\; \alpha' \;\; \text{and} \;\; \tau_2 \;\;
\text{fixed})
\eq
the Hamiltonian reduces to a system of decoupled
harmonic oscillators, and the theory can be solved exactly.
In particular, we can quantize the system and determine the mass
spectrum.
At this point it is important to note that since the states we are
interested in considering are BPS, their mass is exact, and can be trusted for all
radii. Thus, even though the theory was solved for the special
limit (\ref{limit}), such states can be studied more generically,
a valuable consequence of the BPS condition.

We should also point out that as long as $w_0=w_9 \, w_{11}\neq0$,
the spectrum of the Hamiltonian is discrete. If one was
considering $R^{10}\times S^1$, there would be flat directions in
the Hamiltonian, causing the membrane to be unstable, and the
spectrum would be continuous.

After having dropped $H_{int}$ one can proceed to the quantization
of $H_0$. The fields can be expanded in terms of creation and annihilation
operators ($a=1,\ldots,8,11 $),
\bea
X^a_{(k,m)}&=&\frac{i}{\sqrt{2}\,\omega_{(k,m)}} \,
[\alpha^a_{(k,m)}+\tilde{\alpha}^a_{(-k,-m)}]\, , \;\;\;\;\;\;
P^{\,a}_{(k,m)}=\frac{1}{\sqrt{2}} \,
[\alpha^a_{(k,m)}-\tilde{\alpha}^a_{(-k,-m)}]\, , \nonumber \\
(X^a_{(k,m)})^\dag &=& X^a_{(-k,-m)}\, , \; \; \; \;(P^{\,a}_{(k,m)})^\dag =P^{\,a}_{(-k,-m)},
\;\;\;\;\omega_{(k,m)}=\text{sign}(k)\, \omega_{km}\, .
\ea
Canonical commutation relations give
\beq
\left[ X^a_{(k,m)},P^b_{(k',m')}\right] = i
\delta_{k+k'}\delta_{m+m'}\delta^{ab},\;\;\;\;
\left[ \alpha^a_{(k,m)},\alpha ^b_{(k',m')} \right] = \omega_{(k,m)}\delta_{k+k'}\delta_{m+m'}\delta^{ab}\nonumber  \\
\eq
and similarly for the $\tilde{\alpha}$'s.
The explicit form for the time-dependent part of $X^a$ becomes
\beq
X^a(\tau,\s,\rho)=x^a+\alpha' p^a \tau + i\sqrt{\frac{\alpha'}{2}}
\sum_{(k,m)\neq(0,0)}\frac{e^{iw_{(k,m)}\tau}}{\omega_{(k,m)}}
\left[\alpha^a_{(k,m)}e^{ik\s+im\rho}+\tilde{\alpha}^a_{(k,m)}e^{-ik\s-im\rho}
\right].
\eq
The quadratic Hamiltonian (\ref{H0}) takes the form
\beq
\alpha' H_0= \frac{1}{2}\alpha'(p_9^2+p_{11}^2+p_i^2)+\frac{R_9^2
w_0^2}{2\alpha'}+{\cal H},
\eq
with $i=1,\ldots,8 $ and where ${\cal H}$ contains the contributions from the oscillators,
\beq
{\cal H} = \frac{1}{2} \sum_{m,k}^\infty \bigl[\alpha^a_{(-k,-m)}\alpha^a_{(k,m)}+
\tilde{\alpha}^a_{(-k,-m)}\tilde{\alpha}^a_{(k,m)}\bigr].
\eq
Letting the momenta in the $X^9,X^{11}$ directions be
$p_9=\frac{l_9}{R_9}$ and $p_{11}=\frac{l_{11}}{R_{11}}$, with $l_9,l_{11}$
integers,
the {\it nine-dimensional} mass operators becomes
\beq
\label{mass}
M^2 =2\, H - p_i^2 =
\frac{l_9^2}{R_9^2} + \frac{l_{11}^2}{R_{11}^2} + \frac{R_9^2 w_0^2}{\alpha'^2}
+ \frac{2}{\alpha'}\,{\cal H}.
\eq
Clearly, from the eleven-dimensional point of view $\frac{l_9}{R_9}$ and $\frac{l_{11}}{R_{11}}$
are just momenta, but they play the role of mass terms in nine
dimensions.
Schwarz \cite{Schwarz:1995du} showed that the
non-oscillating part of the spectrum,
$M_0\equiv \frac{l_9^2}{R_9^2} + \frac{l_{11}^2}{R_{11}^2} + \frac{R_9^2 w_0^2}{\alpha'^2}$,
matched the corresponding spectrum of the (non-oscillating) type IIB $(p,q)$ string.

As in the case of the string, the level matching conditions
for the membrane are obtained from the global contraints
\beq
P^{(\sigma)}=\frac{1}{2\pi\alpha'}\, \int_0^{2\pi} d\sigma \; \p_\sigma X^a \dot{X}^a
\equiv 0, \;\;\;\;\;
P^{(\rho)}=\frac{1}{2\pi\alpha'}\, \int_0^{2\pi} d\rho \; \p_\rho X^a
\dot{X}^a \equiv 0.
\eq
These can be re-written in terms of mode operators in the
following way,
\beq
N^+_\s -N^-_\s = w_9 l_9, \;\;\;\;N^+_\rho -N^-_\rho = w_{11} l_{11},
\eq
where
\bea
N^+_\s &=& \sum_{m=-\infty}^{\infty}\sum_{k=1}^{\infty}\frac{k}{\omega_{km}}
\, \alpha^i_{(-k,-m)}\alpha^i_{(k,m)}, \; \; \;
N^-_\s =\sum_{m=-\infty}^{\infty}\sum_{k=1}^{\infty}\frac{k}{\omega_{km}}
\, \tilde{\alpha}^i_{(-k,-m)}\tilde{\alpha}^i_{(k,m)}, \nonumber \\
N^+_\rho &=& \sum_{m=1}^{\infty}\sum_{k=0}^{\infty}
\frac{m}{\omega_{km}}\,
\bigl[\alpha^a_{(-k,-m)}\alpha^a_{(k,m)}+
\tilde{\alpha}^a_{(-k,m)}\tilde{\alpha}^a_{(k,-m)}\bigr] , \; \; \;
\nonumber \\
N^-_\rho &=& \sum_{m=1}^{\infty}\sum_{k=0}^{\infty}
\frac{m}{\omega_{km}}\,
\bigl[\alpha^a_{(-k,m)}\alpha^a_{(k,-m)}+
\tilde{\alpha}^a_{(-k,-m)}\tilde{\alpha}^a_{(k,m)}\bigr].
\ea

We are interested in considering states that are BPS.
If we want to add a wave to the 2-brane background while preserving
supersymmetry, we must align it \cite{Russo:1996if} along the momentum
direction. Furthermore, for a BPS state one can have only
right-moving oscillations.
For the special case $l_{11}=0$, these conditions imply
that the oscillations are only along the $\sigma$
direction, $N_\rho^\pm=0$, and that there are no left-moving oscillators, $N_\s^-=0$.
The condition $N_\rho^+=0$ implies that these states are built by
applying $\alpha^i_{(-k,0)}$ to the vacuum, which gives $\omega_{km}\rightarrow w_{11} k$.
Thus, one finds
\beq
{\cal H}=w_{11}\,N_\s^+=w_0 l_9,
\eq
yielding
\beq
M^2_{BPS}=\frac{l_9^2}{R_9^2}+ \frac{R_9^2 w_0^2}{\alpha'^2}
+ \frac{2}{\alpha'}\,{w_0 l_9}=\Bigl(\frac{l_9}{R_9}+4\pi^2 T_3\,w_0\,R_9 \, R_{11}\Bigr)^2.
\eq
For a general state with $l_9,l_{11}\neq 0$, the constraints become \cite{Russo:1996if}
\beq
\label{constraints}
N_\s^+=l_9 w_9, \;\;\;\;\; N_\rho^+=l_{11} w_{11},
\eq
giving
\beq
{\cal H}=w_0 \sqrt{l_9^2+l_{11}^2\tau_2^2}.
\eq
Thus, we obtain the more general mass formula
\beq
\label{GeneralSpectrum}
M^2_{BPS}=\Bigl(\sqrt{\frac{l_9^2}{R_9^2}+\frac{l_{11}^2}{R_{11}^2}}
+ 4\pi^2 T_3 w_0 R_9 R_{11}\Bigr)^2,
\eq
matching the spectrum of the oscillating $(p,q)$ string \cite{Russo:1996if}, as
anticipated by Schwarz.
One can rewrite the mass spectrum in a more convenient form,
\beq
\label{massR11mod}
M^2_{BPS}=m_P^2 \Bigl(\sqrt{\frac{l_9^2}{R_9^2}+\frac{l_{11}^2}{R_{11}^2}}
+w_0 R_9 R_{11}\Bigr)^2,
\eq
where $m_P$ is the Planck mass, and the radii are now dimensionless.
We are particularly interested in the case $l_9=0$, yielding
\beq
\label{massR11}
M^2_{BPS}=m_P^2 \Bigl(\frac{l_{11}}{R_{11}}+w_0 \,R_9
\,R_{11}\Bigr)^2.
\eq
We would like to point out that, since the particle numbers $N_\s^+,
N_\rho^+$ must be positive, the constraints (\ref{constraints})  imply that $w_9,
l_9$ must have the same sign (and similarly for $w_{11},l_{11}$).
Thus, the states described by (\ref{GeneralSpectrum}) are always
massive \footnote{We would like to thank K. Hori for discussions regarding this point.}
for non-trivial quantum numbers.

As we will show in Section \ref{ClTrapping}, such states can
play an interesting role on cosmological evolution.
However, we are particularly interested in the cosmological consequences of
states having a mass of the form
\beq
\label{ourmass}
M^2\sim m_P^2 \Bigl(\frac{|l_{11}|}{R_{11}}-|w_0| R_9 \,R_{11}
\Bigr)^2,
\eq
which can become {\it massless}, and signal an enhancement of symmetry at $R_9\sim R_{11}\sim 1$ (for the
case $|l_{11}|=|w_0|)$.

Such states do {\it not} appear in the supermembrane spectrum derived in
\cite{Russo:1996if}, however one may expect them to be present in the spectrum of M-theory,
possibly in heterotic M-theory.  In fact,
heterotic string theory contains
states which become massless at enhanced symmetry points, much in the same way as
in the bosonic string case.
The mass spectrum of the heterotic string with a compact dimension
of radius $R$ is of the form
\beq
\frac{1}{4}M_{\,h}^{\,2}=\Bigl(\frac{p_L^{\,2}}{2}+N_L-1 \Bigr) + \Bigl(\frac{p_R^{\,2}}{2}+N_R-c_R \Bigr)
\eq
where $p_{L,R}=n/R \pm w \, R$, and
for clarity we have separated the contributions from the
left-moving and right-moving sides.
For the NS sector $c_R=\frac{1}{2}$, while for the R sector
$c_R=0$. Thus, the presence of the zero-point energy allows for
states of the form
\beq
\label{radion}
M^{\,2} \sim m_s^2\Bigl(\frac{1}{R}-R\Bigr)^2
\eq
with $R$ the radius of any one of the six compact dimensions, i.e.
the {\it radion}.
Thus, the heterotic string shows enhancement of symmetry as $R$ approaches the
self-dual radius.
At the moment we are not aware of any explicit
construction of states of the form (\ref{ourmass}), where the
pivotal role is played by the {\it eleven-dimensional radius}, i.e. the
dilaton. However, in analogy with what happens in the case of the
radion, eq. (\ref{radion}), we expect
that states of the form (\ref{ourmass})
should be found in M-theory on $S^1/\mathbb{Z}_2$.

The lack of such configurations in the derivation of \cite{Russo:1996if}
is due to the absence of the zero-point energy in
the supermembrane quantization, which is believed to be cancelled
by fermionic contributions \cite{Duff:1987cs}.
It is also consistent with the fact that such (potentially
massless) states are not found in type IIA or IIB string theory, which are
obtained from dimensional reduction of 11D M-theory.
However, we should mention that supermembrane quantization is still not
well understood, and that a full quantum theory of the membrane has
proven very difficult to obtain.
If, in fact, the zero-point energy were not entirely
cancelled by fermions, or if supersymmetry was broken, one
could still be able to find states of the form
(\ref{ourmass}).
Another approach would be to relate
known instances of enhanced symmetry to M-theory, through
appropriate use of the available dualities, dimensional reduction (and oxidation).
Lastly, we mention that considering the supermembrane on backgrounds other than the torus may also offer a resolution,
noting for example that Type II strings on K3 (dual to heterotic on $T^6$) also possess enhanced symmetry.

Finding whether enhanced symmetry states of the form (\ref{ourmass}) can be present is an interesting problem in its own
right, which we would like to consider in the future.
However, for the purpose of this paper, we will simply postulate that
these states should exist and restrict ourselves to
the study of their cosmological consequences.

\section{Classical Approach to Trapping \label{ClTrapping}}

In the last section we found a set of dyonic solutions
whose mass depends explicitly on the dilaton or, from the $11$D
perspective, $R_{11}$:
\beq
\label{heavymass}
M^{\,2}=m_p^2\Bigl(\frac{l_{11}}{R_{11}}+w_0 R_9 R_{11}\Bigr)^2.
\eq
From now on we will set $m_p=1$, and restore explicit $m_p$
dependence only when needed.
We will now consider the effect of a gas of these dyonic states
treated as {\it classical} sources for the vacuum field equations
of $11$D SUGRA.  The low-energy effective action for the bosonic
degrees of freedom is
\beq \label{11daction}
S=\frac{1}{2\kappa_{11}^2}\int d^{11}x \sqrt{g}
\left[R-\frac{1}{48} G_4^2\right]+\frac{1}{6} \int A_3 \wedge G_4
\wedge G_4,
\eq
where $G_4$ is the anti-symmetric four-form flux
$G_4=dA_3$, and $\kappa_{11}^2=8\pi G_{11}$, with $G_{11}$ the
eleven-dimensional Newton constant. Anticipating a universe with
three large spatial dimensions we consider the following ansatz
for the metric and four-form
\bea \label{ansatz}
ds^2&=&-e^{2A}d\eta^2+e^{2B}d\vec{x}\,^2+e^{2C}d\vec{y}\,^2+e^{2D}dz^2,\\
G_4&=& h \, dx^1 \wedge dx^2 \wedge dx^3 \wedge dz,
\ea
where $\vec{x}=(x_1,x_2,x_3), \, \vec{y}=\{y_m\}, \, m=4,\ldots,9$, and
$z$ denotes the coordinate of the eleventh dimension. Notice also that
we have introduced exponential scale factors $R_m=e^{\,C}$ and
$R_{11}=e^{\,D}$. The functions $A,B,C$ and $D$ depend only on
$\eta$, and $h$ is a constant. For this choice of flux both the
equation of motion and the Bianchi identity for the gauge field
are trivially satisfied.  The remaining equations of motion follow
from varying the action with respect to the full metric $g_{MN}$,
with $M,N=(0,1,\ldots,9,11)$, \beq
R_{MN}-\frac{1}{2}g_{MN}R=\kappa_{11}^2 \left(
\frac{1}{12}F_{MOPQ}F_N^{\; \;
OPQ}-\frac{1}{96}g_{MN}F^2+T_{MN}^{\; \text{sources}} \right), \eq
where the dyon sources (\ref{heavymass})
are included through their stress tensor $T_{MN}^{\; \text{sources}}$.

Working in the ideal gas approximation the energy density of these
sources can be found from the mass, and is given by
\beq
\rho=\frac{M}{V}=e^{-3B-6C-D} \left( l_{11} e^{-D}+w_0 e^{C} e^{D}
\right).
\eq
The pressures are then given by
\beq \label{pressures}
p_{\,3}=-\frac{1}{3}\frac{\partial \rho}{\partial B}-\rho, \;\;\;\;
p_{\,6}=-\frac{1}{6}\frac{\partial \rho}{\partial C}-\rho, \;\;\;\;
p_{\,11}=-\frac{\partial \rho}{\partial D}-\rho,
\eq
yielding for the eleven-dimensional pressure
\beq
p_{\,11}=e^{-3B-6C-D} \left( l_{11} e^{-D}-w_0 e^{C} e^{D}
\right).
\eq
If we take the six dimensions to be at the self-dual radius (i.e. $R_m=1$, or $C=0$)
we find that the above pressure will act to stabilize $R_{11}$, corresponding
to driving $D \rightarrow 0$.
There, the pressure vanishes (for the case $l_{11}=w_0$), and the energy
density has a local minimum.
We should note that in this analysis the flux would not play a major role, since we are
interested in solutions with $B(\eta)$ growing large, and one
finds that the flux
\footnote{However, the flux will be vital at early times (small $B$) \cite{Friess:2004zk}.}
scales as $\sim h^2e^{-6B}$.

Such classical attractors have been studied in many contexts throughout the literature.
Examples include gases of wrapped strings and branes
(see \cite{Battefeld:2005av,Brandenberger:2005nz,Brandenberger:2005fb, Patil:2005nm}
and references within), blackhole
attractors \cite{Kaloper:2004yj}, D-brane systems \cite{Abel:2005jx}, and
cosmologies involving conifold and flop
transitions, e.g., \cite{Mohaupt:2005pa}.  Despite these elegant
ideas for stabilizing moduli, a detailed analysis shows that initial conditions
play a crucial role.  In the example we have here, one finds that
stabilization of $R_{11}$ can be quite generic for fixed $C$.
However, as noticed in \cite{Berndsen:2005qq},
attempting to stabilize all dimensions simultaneously proves
difficult and requires extreme fine-tuning.
But perhaps a more serious objection is the {\it validity} of this classical approach.
The states (\ref{heavymass}) that we have considered thus far are very heavy for generic values
of the moduli.  In fact, even at the stabilization point, the energy of these states
is typically near the Planck scale.
Thus, the low-energy effective theory that provided us with the equations of motion is no
longer the proper
framework, given the inclusion of massive states. Furthermore, one
cannot arbitrarily add only some, but not all, of the massive
states of the strings or branes.

We will turn to a more self-consistent approach in the next section, but it is important
to stress that, although the classical approach of this section was naive,
it is still interesting to see how the inclusion of truly {\em stringy} states
can affect the dynamics.
That is, low-energy SUGRA is {\it not} string theory, and the addition of the
membrane winding and oscillator contributions can be considered a first correction
toward the UV completion of the theory.  Thus, a better
understanding of the full string (or membrane) spectrum is a first
step towards addressing the string vacuum problem and exploring
string phenomenology.
However, we do support the viewpoint that this approach is not
truly consistent, and whereas in the next section we will find a self-consistent way to
include {\em stringy} states in the low-energy theory.

\section{Membrane Production in Time-Dependent Backgrounds \label{QmTrapping}}

In this section we consider the corrections which may result from the dynamics
of moduli in the low-energy effective action of string (M-)theory, namely SUGRA.
As we have seen, string models contain states whose masses depend explicitly on
massless moduli.
If the moduli are time-dependent, the masses of such string
states change, and can even vanish.
Of course, this mechanism is not specific to string theory;
the Higgs is a field theory example of a modulus which controls
particle masses, however the crucial difference here is the target space of the string is space-time itself.

Typically the states become light as the moduli approach
locations where symmetries play a special role (ESPs).
Near such points, the states can be quantum produced, and must
therefore be incorporated into the effective field
theory description.
As an example of this {\em stringy} Higgs mechanism, consider the case of a D-brane wrapping a collapsing cycle of the extra dimensions.
It was realized some time ago \cite{Strominger:1995cz}, that by including additional light states resulting from the
collapse of a wrapped D-brane on a shrinking cycle of the Calabi-Yau, the naive singularities of the conifold geometry could be resolved.
The singularity in the moduli space was realized as an artifact of integrating out degrees of freedom that were no longer massive and beyond the cutoff
of the effective theory.  We see that dynamics can play a vital role when consider which degrees of freedom to include in the effective theory,
since the mass scale can be dynamical given the evolution of background fields (in this case the collapse of the extra dimensions and the brane).

One can identify several corrections to the moduli space approximation
resulting from the inclusion of the new light degrees of freedom.
Although, in this paper we will concentrate on on-shell production
neglecting other types of corrections, which we will mention
briefly in the conclusions.
As we will see, particle production will be suppressed generically,
but it will be non-zero and significant for the enhanced symmetry
states mentioned at the end of Section \ref{MthStates}.
We will then study the effects of the resulting backreaction on
cosmological evolution, and show that it can lead to trapping of
the moduli under consideration.
Such a trapping mechanism has been previously studied in \cite{Watson:2004aq,Kofman:2004yc}.

\subsection{Time-Dependent Backgrounds}

Let us begin by considering the low energy effective theory for M-theory, i.e. $11$D SUGRA.
The effective theory for the massless bosonic fields is given by the action (\ref{11daction}).
A class of solutions was found in \cite{Lidsey:1999mc}
\bea
\label{FullScaleFactors}
ds^2&=&-e^{2A}d\eta^2+e^{2B}d\vec{x}\,^2+e^{2C}d\vec{y}\,^2+e^{2D}dz^2, \nonumber \\
G_4&=& h \, dx^1 \wedge dx^2 \wedge dx^3 \wedge dz,
\eea
with the scale factors given by
\bea
A&=&3B+D \nonumber \\
B&=&B_0+\frac{1}{3}\log{\sec{h\eta}}+
\frac{q}{2}\log{\frac{1+\sin{h\eta}}{\cos{h\eta}}} \nonumber \\
C&=&C_0+\frac{1}{6}\log{\cos{h\eta}} \nonumber \\
D&=&D_0-\frac{1}{3}\log{\cos{h\eta}}+\frac{1-3q^2}{6q}
\log{\frac{1+\sin{h\eta}}{\cos{h\eta}}},
\ea
where $B_0,C_0,D_0 \; \text{and} \; |q| \leq 1/\sqrt{3}$ are real constants and $-\frac{\pi}{2h}  \leq \eta  \leq \frac{\pi}{2h}$.
Notice that in the limit of no flux, $h=0$, we recover flat space and  $-\infty \leq \eta \leq \infty$.
We also note that, although this space-time contains a singularity (located at
$-\frac{\pi}{2h}$ for $q>0$),
we will only be interested in the evolution away from the singularity
\footnote{In \cite{Friess:2004zk}, winding mode creation in this background was studied
in order to try to resolve the cosmological singularity.
In this paper, however, we will focus on production away from the singularity.}.

We now want to consider the production of membrane states
of the type discussed in Section \ref{MthStates},
having an effective mass
which depends on the radii of the compact dimensions.
The first question one may ask is why we expect any production at all, since
at low energy scales production of string and membrane states should be strongly
suppressed.
For the moment let us push forward, keeping this important question in mind.
Consider the quantum equation of motion for a string/membrane state labelled by $\chi$.
As discussed in \cite{Lawrence:1995ct,Gubser:2003vk} the string constraint equations amount to a
wave equation for $\chi$,
\be \label{chieom}
{\chi}^{\prime \prime}_k+\omega_k^2 \chi_k=0,
\ee
where a prime denotes the derivative with respect to $\eta$, and $\chi_k$
are the Fourier modes of $\chi$.
We have removed the friction term by a field redefinition $\chi \rightarrow \chi e^{-3C}$.
For simplicity, we will rescale time $\eta \rightarrow \eta h^{-1}$ so that
$-\frac{\pi}{2}< \eta < \frac{\pi}{2}$.
The time dependent frequency is then given by
\be \label{omega}
\omega^2=e^{6B+2D}h^{-2}\left(  \frac{k_3^2}{e^{2B}} + \frac{k_6^2}{e^{2C}} + \frac{k_{11}^2}{e^{2D}} +m^2(\eta) +\xi R -e^{-6B-2D}h^2(9C^{\prime \, 2}
+3C^{\prime \prime}) \right),
\ee
where $R$ is the $11$D Ricci scalar, $\xi$ is the coupling of $\chi$ to the
space-time background, and $m^2$ represents the contributions to the mass coming
from the winding and oscillations of the membrane.
Also note that $k_3, k_6$ and $k_{11}$ denote {\it comoving}
momentum.
It will be more convenient at times to think in terms of the nine-dimensional mass
\bea \label{masseff}
m^2_{{eff}}&=&\left( k^2_{11}e^{-2D} +m_w^2+m_{\text{osc}}^2 \right) \nonumber \\
&=&m_p^2 \left(  l_{11}e^{-D}-\omega_0 e^D e^C \right)^2,
\eea
since it is when this mass vanishes that we expect significant production.
Choosing $B_0=C_0=D_0=0$ for simplicity, these states
will become massless (for the case $l_{11}=w_0$) at $\eta=0$, where $D=C=0$.
Given the time-dependent frequency (\ref{omega}), with $l_{11}$ and $w_0$
left arbitrary, we are now ready to consider the production of the
membrane states.

\subsection{The Steepest Decent Method \label{sdm}}

Before considering the case of interest, let us review the
standard method for calculating particle production of states with
a time-dependent frequency (see e.g. \cite{Birrell:1982ix}). A
formal solution to the mode equation (\ref{chieom}) is given by
\beq \chi_k = \frac{\alpha_k}{\sqrt{2\omega_k}}e^{-i\int \omega_k
d\eta}+\frac{\beta_k}{\sqrt{2\omega_k}}e^{i\int \omega_k d\eta}.
\eq The normalization condition for the scalar field is then
$|\alpha_k(\eta)|^2-|\beta_k(\eta)|^2=1$, which can be used to
write the equation of motion as \bea \alpha_k' &=&
\frac{\omega_k'}{2\omega_k} e^{2i\int
\omega_k(\eta')d\eta'}\beta_k \nonumber \\ \beta_k' &=&
\frac{\omega_k'}{2\omega_k} e^{-2i\int
\omega_k(\eta')d\eta'}\alpha_k. \ea Initially we start near the
adiabatic vacuum and we have $\beta_k<<1$ and $\alpha_k \sim 1$ so
that \beq \label{beta} \beta_k \sim \int d\eta
\frac{\omega_k'}{2\omega_k} e^{-2i\int^\eta
\omega_k(\eta')d\eta'}, \eq to leading order.  Since $|\beta|^2$
gives the number of particles produced, we see that production is
suppressed as long as $\omega^\prime \ll \omega$. Conventionally,
one defines the dimensionless non-adiabatic parameter
${\omega^\prime}/{\omega^2}$, indicating that particle production
becomes significant when
\be
\frac{\omega^\prime}{\omega^2} \sim 1.
\ee
In order to calculate the total amount of particles produced we need to estimate the integral in
(\ref{beta}).  A method for approximating this integral was found
in \cite{Chung:1998bt}, which we will summarize below.

We will be interested in $\omega_k$ of the form $\omega_k=\sqrt{k^2+m^2 f(\eta)}$,
where we absorb any time dependence of the background into $f(\eta)$.
The poles of the integrand in (\ref{beta}) are also
branch points in the complex $\eta$ plane.  Let us
assume for simplicity that $\omega_k$ has a single pole at
$\eta^\star$.
Then, near the branch point, the integral in the exponent of (\ref{beta})
can be expanded in the following way:
\beq
\int_{\eta_{in}}^\eta \omega_k(\eta') d\eta' =\int_{\eta_{in}}^\eta \omega_k(\eta')
d\eta' + \frac{2M}{3}\sqrt{f'(\eta^\star)}(\eta-\eta^\star)^{3/2},
\eq
keeping the leading term in the expansion of $f(\eta)$ about $\eta^\star$.
We then find
\beq
\beta_k = \Bigl( \frac{1}{4}\int_{f^\star}\frac{d\delta}{\delta}e^{\frac{-4iM}{3}
\sqrt{f'(\eta^\star)}\delta^{3/2}} \Bigr) \,
e^{-2i\int_{\eta_{in}}^{\eta^\star} \omega_k(\eta')d\eta'},
\eq
where the $f^\star$ denotes the steepest descent contour.
The factor in front of the exponential can be shown to equal
$\frac{i\pi}{3}$, giving us
\beq
\beta_k = \frac{i\pi}{3} e^{-2i\int_{\eta_{in}}^{r} \omega_k(\eta')d\eta'} \,
e^{-2i\int_{r}^{\eta^\star} \omega_k(\eta')d\eta'},
\eq
where $\eta^\star=r-i\mu$.
The first integral in the expression
above is real, making the first exponential a pure phase.
The second integral can be approximated by
$\int_{r}^{\eta^\star} \omega_k(\eta')d\eta' \sim i \pi \mu
\omega_k(r)$, yielding
\beq \label{betaform}
|\beta_k|^2 = \biggl(\frac{\pi}{3}\biggr)^2 e^{-\pi\mu\omega_k(r)}.
\eq

Thus, to calculate the leading contribution to particle production
it is sufficient to identify the real and imaginary contributions
to the zeros of $\omega_k$. Because of the choice of integration
contour, we should note that we are only interested in the zeros which are in the
lower half plane.

\subsection{Membrane Production}

We now want to determine which conditions will lead to significant production
in the chosen background (\ref{FullScaleFactors}).  The
adiabatic approximation is valid as long as ${\omega^\prime}/{\omega^{2}}<1$,
and as long as this condition holds production will be insignificant.
Thus, the first step is to examine when the non-adiabaticity is appreciable for our frequency
\be \label{ourfreq}
\omega^2_k=e^{6B+2D}h^{-2}\left(  \frac{k_3^2}{e^{2B}}  +m_{eff}^2(\eta)
+\xi R -e^{-6B-2D}h^2(9C^{\prime \, 2}+3C^{\prime \prime}) \right),
\ee
where we will neglect any momentum in the extra dimensions (i.e. we take $k_6=0$),
but keep $k_{11}$ non-zero.

One obvious location where non-adiabaticity becomes large is near the cosmological
singularity ($\eta=-\pi/2$), which was the case examined in \cite{Friess:2004zk}.
Here, however, we are not interested in the region near the
singularity. Thus, as long as we are not too close to the singular
region, we can still construct the appropriate in-vacuum
\footnote{This is analogous to the calculation of production at the end of inflation,
where the initial big-bang singularity is irrelevant to the calculation.
See \cite{Birrell:1982ix} for a detailed discussion on the method of constructing
an asymptotic, adiabatic vacuum in such singular space-times.}.
Let us now examine the behavior of the non-adiabatic parameter away from
the initial singularity.  We expect that away from the values $C=D=0$ the mass term
should dominate over all other terms in the frequency (\ref{ourfreq}).
In this limit, $$\frac{\omega^\prime}{\omega^2} \approx \frac{m_{eff}^\prime}{m_{eff}^2},$$
and we expect the non-adiabaticity to be peaked near $m_{eff}\approx 0$, which is where
the states become massless.

In Figures \ref{fig1}, \ref{fig2}, and \ref{fig3}, we present the behavior of the
{\it exact} non-adiabatic parameter for various values of the background and momenta.
In all cases we find that the production is sharply peaked around $\eta \approx 0$.
As noted in \cite{Kofman:2004yc} (and references within), this means that production can
be treated as an instantaneous event, making the calculation of production
and its backreaction much more tractable.  We see from the figures that
away from $\eta \approx 0$ the adiabatic vacuum is an excellent approximation,
and that well-defined in and out regions do indeed exist.

\begin{figure}
\includegraphics[width=10.5cm]{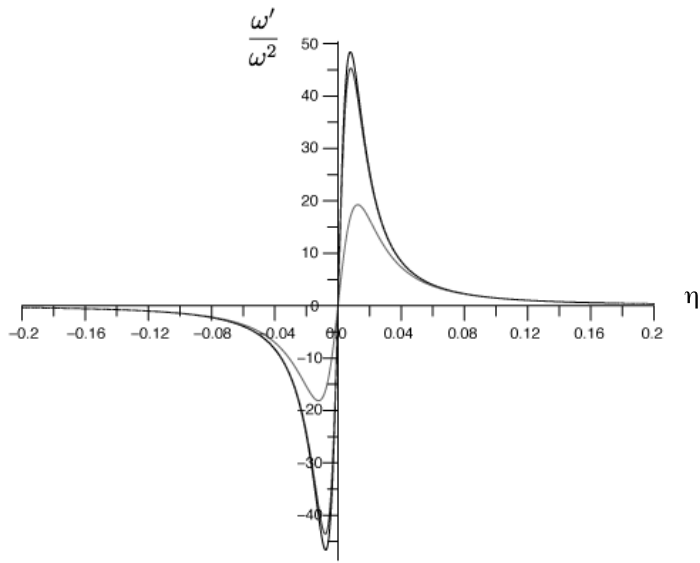}
\caption{\label{fig1} The non-abiabatic parameter ${\omega^\prime}/{\omega^{2}}$
for different values of $\xi$, with
$\xi=0$ (top curve), conformal coupling $\xi=9/40$ (middle curve), and $\xi=5$ (bottom curve).
As $\xi$ increases, non-adiabaticity is suppressed.
Note that the non-adiabticity is
concentrated near $\eta \approx 0$ for all choices of $\xi$.}
\end{figure}
\begin{figure}[!]
\includegraphics[width=9.5cm]{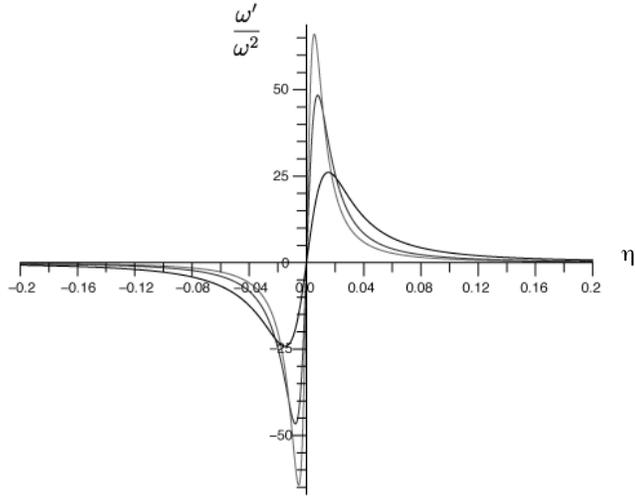}
\caption{\label{fig2} The non-adiabatic parameter ${\omega^\prime}/{\omega^{2}}$ for different
values of the flux $h=1/150$ (top curve), $h=1/100$ (middle curve), and $h=1/50$ (bottom curve).}
\end{figure}
\begin{figure}[!]
\includegraphics[width=9.5cm]{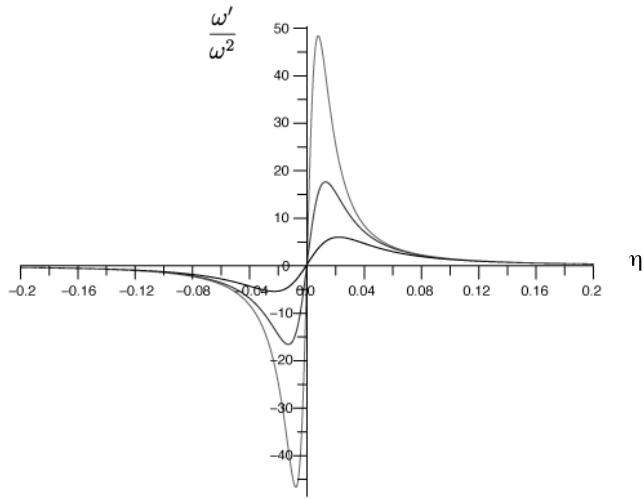}
\caption{\label{fig3} The non-adiabatic parameter ${\omega^\prime}/{\omega^{2}}$ for different values of the
momentum $k_3=1/500$ (top line), $k_3=1/100$ (middle line), and $k_3=1/50$ (bottom line).}
\end{figure}

One point of concern is the effect of the time-dependent geometry on the
effective mass  $$m^2_{total}=m_{eff}^2+\xi R + \text{gravity terms}$$ through the
last three terms in (\ref{ourfreq}).
As seen in Fig. \ref{fig1}, the terms including the coupling to the geometry $\xi$,
have the effect of suppressing the amount of production, but this suppression is
negligible for reasonable values of the coupling.  In particular, the cases of minimal
($\xi=0$) and conformal $(\xi=9/40)$ coupling have little or no effect.  This can be understood
by examining the Ricci curvature which, for small flux, is negligible away from the
singular region (i.e., $R \sim h^2$).  The requirement of small flux can be seen
from Fig. 2, where we find that production will be suppressed unless $h<m_p$.
Fig. 3 shows a similar result for the momenta, $k_3 < m_p$.
Let us now consider the production from a more quantitative perspective.

We have seen that the non-adiabaticity is focused near $\eta \approx 0$.  Using this
and the effective mass (\ref{masseff}), we can expand the frequency (\ref{ourfreq})
for small $\eta$
\be \label{approxomega}
\omega_k^2(\eta)=f_0+f_1 \eta+f_2 \eta^2,
\ee
where, after restoring the explicit dependence on $m_p$ ,
\bea
f_0&=& \frac{k_3^2+a_0m_p^2+\frac{1}{2} h^2}{h^2}, \;\;\;\;\;\;\;
f_1 = \frac{a_1 k_3^2+a_2 m_p^2}{h^2}, \;\;\;\;\;\;\;
f_2 = \frac{a_3k_3^2+a_4m_p^2+\frac{1}{4} h^2}{h^2}, \nonumber \\
a_0&=&(w_0-l_{11})^2,
\;\;\;\;a_1=\frac{1+3q^2}{3q}, \;\;\;\;
a_2=\frac{2w_0(w_0-l_{11})+3q^2(w_0^2-4w_0l_{11}+3l_{11}^2)}{3q} \nonumber \\
a_3&=&\frac{1+24q^2+9q^4}{18q^2}, \;\;
a_4=\frac{2w(w-l)+3q^2(13w^2-23wl+6l^2)+q^4(9w^2-72wl+91l^2)
}{18 q^2}.\nonumber
\ea
Following the method outlined in Section \ref{sdm}, we proceed by finding the zeros of (\ref{approxomega}).
The frequency vanishes at
\beq
\eta^*=\frac{-f_1\pm\sqrt{f_1^{\, 2}-4\, f_0 \, f_2}}{2f_2}.
\eq
Remembering that we are interested in the zeros in the lower
half plane, we choose
\beq
\eta^\star=-\frac{f_1}{2f_2}-i\frac{\sqrt{4f_0
f_2-f_1^2}}{2f_2}\equiv r-i\mu,
\eq
where we have $4f_0 f_2-f_1^2>0$.
The amount of particle production for a given mode is then given by (\ref{betaform}),
and is found to be
\beq
|\beta_k|^2 = \biggr(\frac{\pi}{3}\biggr)^2 e^{-\pi\frac{4f_0
f_2-f_1^2}{4 \, f_2^{3/2}}}.
\eq
Moreover, using the fact that $m_p\gg k_3,h$, we find
\beq \label{betaeqn}
|\beta_k|^2 = \biggr(\frac{\pi}{3}\biggr)^2 e^{-\frac{A_1\,m_p}{h}}
e^{-\frac{A_2\,k_3^2+A_3\,h^2}{m_p h}+{\cal
O}(\frac{1}{m_p^2})},
\eq
where $A_1, A_2, A_3$ are constants that depend on $w_0, l_{11},
q$.

Given this expression we can now return to the initial concern that string and
membrane particle production should suffer Planck scale suppression
at low energies.  This can be seen by the suppression factor $e^{-\frac{A_1\,m_p}{h}}$
appearing in (\ref{betaeqn}), which is related to the usual suppression
due to the Hagedorn density of states (see e.g. \cite{Gubser:2003vk}).
However, in this case we find that $A_1 \propto (l_{11}-w_0)^3$, which vanishes
when $l_{11}=w_0$, consistent with the condition for having
massless states\footnote{This is the motivation for wanting states
having a mass of the form (\ref{ourmass}).}
 at the self-dual radius $D=0$.  Thus, the enhanced symmetry results in
additional light states which will be copiously produced.
In fact, we find significant particle production when $l_{11}=w_0$,
given by
\beq \label{betaeqna}
|\beta_k|^2 = \biggr(\frac{\pi}{3}\biggr)^2 e^{-\frac{3\pi}{2\,m_p\,h}
\bigl(\frac{h^2+2 k_3^2}{l_{11}\gamma}
\bigr)},
\eq
where $\gamma\equiv \frac{1-3q^2}{q}$.
Furthermore, we will choose $l_{11}=w_0=1$, since states
with higher winding and momentum numbers would decay to this configuration.
We can immediately see from (\ref{betaeqna}) the same behavior that we found for the non-adiabatic parameter in the previous section.
Namely, we see that for $h, k_3 < m_p$ production will result and above the Planck scale production will cease.  This behavior agrees with the numerical studies
of the non-adiabatic parameter, which can be seen in Fig. \ref{fig2} and Fig. \ref{fig3}.

From (\ref{betaeqna}) we can calculate the eleven-dimensional number
density at the creation time $\eta^*$ by summing over all modes
\beq \label{densities}
n_\chi (\eta^*)
=\frac{1}{V_0^{(7)}} \int_0^\infty \frac{d^3k_3}{(2\pi)^3}
\; |\beta_k|^2,
\eq
$V_0^{(7)}$ is the dimensionful volume of the extra dimensions
at the time the particles were produced \footnote{We
note that in the calculation of the 11D densities there is an
implicit integration over the momenta of the extra dimensions (and the Kaluza-Klein modes).
However, we have chosen to work with the $l_{11}=1$ states
and any additional momentum (e.g. $k_6$) is taken to vanish.}.
The energy density of created particles is given by
\beq \label{edense}
\rho= \frac{1}{V_0^{(7)}} \int_0^\infty \frac{d^3k_3}{(2\pi)^3} \;  \omega_k
|\beta_k|^2.
\eq
Finally, the pressure can be found from the energy density as in (\ref{pressures}),
and the effective stress energy tensor of the produced membrane states is then
\be \label{st}
T^{M}_{\; \; N}=diag \left( -\rho, p_3, p_6, p_{11} \right).
\ee

\section{Cosmology, Backreaction, and Moduli trapping \label{cosmo}}

We now want to consider the effect of the states produced in the last section
on the evolution
of the background (\ref{FullScaleFactors}).
One possible way to include the effects of such states would be
to study the evolution of
linear perturbations about the background (\ref{FullScaleFactors}).
However, one finds that this approach is not adequate, since the presence of the states
alters the evolution drastically, making the effect of order one.
Instead, we consider the new metric
\be \label{newmetric}
ds^2=-dt^2+a(t)^2 dx_3^2 + b(t)^2 dx_6^2+R_{11}^2(t) dx_{11}^2,
\ee
where we will work in synchronous gauge, and we have assumed that the perturbed metric will
respect the symmetries (topology) of the original metric\footnote{
Note that for simplicity we will take six of the radii to be equal (i.e., $R_m=b$
with $m=4,5,6,7,8,9$).
This could easily be generalized without changing our conclusions.}.
Introducing the Hubble parameters $H_3=\frac{\dot{a}}{a}, H_6=\frac{\dot{b}}{b}$ and
$H_{11}=\frac{\dot{R}_{11}}{R_{11}}$, the equations of motion can be written as
\be \label{theeom}
\dot{H}_i=H_i \sum_{j} H_j + 8\pi G_{11} \left( p_i +\frac{1}{10}
\left(  \rho- \sum_{j} p_j \right) \right),
\ee
with $i,j=1 \ldots 9,11$ running over all spatial dimensions.  We have again taken
the matter sources
to be in the form of a perfect fluid characterized by their energy density and pressures,
and having a stress tensor given by
\be \label{sth}
T_{MN}=diag \left( \rho, a^2 p_3, b^2 p_6, R_{11}^2 p_{11} \right).
\ee

The constraint equation
\be
\sum_{i<j} H_i H_j =8\pi G_{11} \rho
\ee
and the equations motion allow one to obtain the continuity equation
\be
\dot{\rho}=-\sum_i \left( \rho + p_i \right) H_i.
\ee
In general these equations are very difficult to solve.
However, in the absence of matter (i.e., $\rho=p_i=0$) one finds the well-known
Kasner solutions,
\bea
a(t)\sim t^{c_1}, \;\;\;\;\;\;  b(t)\sim t^{c_2}, \;\;\;\;\;\;\;  R_{11}(t)\sim t^{c_3},
\nonumber \\
\text{with} \;\;\;\sum_i c_i = \sum_i (c_i)^2=1.
\eea
Thus, in a universe initially filled with radiation or matter the expansion will
dilute the sources,
and we expect the late-time behavior to approach that of the Kasner solution.

We want to include the effect of the produced states on the background
(\ref{FullScaleFactors}).
As we found in the last section we can include the states through their energy
density (\ref{edense}).
It is important to note that number density is only well defined in the adiabatic
out region.
That is, near the region of non-adiabaticity, particle number (and thus energy density)
can fluctuate.
However, following \cite{Kofman:2004yc} we will assume that after
the first pass through the production point no further production occurs, and the particle
number remains fixed.
This approximation is certainly acceptable, since the exact treatment would merely increase
the number of particles, enhancing the trapping mechanism we will discuss.

Perhaps a more serious drawback is the validity of the SUGRA approximation near the
production point.
For the background we have considered, production will occur near the Planck scale at the
center of the M-theory moduli space.
From the ten-dimensional point of view, this will be a fixed point of S-duality where a
perturbative description of the theory is not available at this time.
However, we can trust our equations away from the non-adiabatic region, and we are primarily
interested in the attractive behavior towards this region.
Thus, we will consider the effects coming from production, remembering
that further corrections may become important near the non-adiabatic region.
In the worst case, we will see that the corrections we consider lead us naturally to the
non-adiabatic region, where a further knowledge of M-theory dynamics is needed.

Using the expression (\ref{betaeqna}) in (\ref{densities}) we find for the number density
at the time of creation $\eta^{*}$
\bea
n_\chi(\eta^{*}) &=& \frac{1}{\vol}
\int_0^\infty \frac{d^3k}{(2\pi)^3} \; \left( \frac{\pi}{3} \right)^2
e^{ -\frac{3\pi}{h\gamma m_p} \left( k_3^2+ \frac{1}{2}h^2 \right) } \\
&=& \frac{\sqrt{3}\gamma^{3/2}}{648 \, \pi}  \left(m_p \, h\right)^{5}
e^{-\frac{3\pi h}{2 \gamma m_p}},
\eea
where we used $V_0^{(7)}=(h\,m_p)^{-7/2}$.
This is the number density of particles resulting from production
at the time when the scale factors pass near the region of non-adiabaticity $D=C=0$.
In terms of the new metric (\ref{newmetric}) the creation time
will be denoted by $t_0$, and the production point corresponds to $R_{11}=b=1$.
Away from the non-adiabatic region the number of particles will remain
constant, and the density in the new metric is given by
\beq
n_\chi(t)=\left( \frac{a^3(t_0) \, b^6(t_0) \, R_{11}(t_0)}{a^3(t)\, b^6(t)\, R_{11}(t)}
\right)n_\chi(t_0)\equiv \frac{V(t_0)}{V(t)}n_\chi(t_0).
\eq

At this point it is convenient to return to setting
$m_p=1$; we will reinstate the explicit dependence on the Planck
mass when necessary.
The energy density at time $t_0$ is given by
\beq
\rho(t_0) =\frac{\pi^2}{9 \vol} \int_0^\infty \frac{d^3k}{(2\pi)^3} \;
\omega_k \, e^{ -\frac{3\pi}{h \gamma} \left( k_3^2+ \frac{1}{2}h^2 \right) }.
\eq
The frequency $\omega_k$ evaluated in the new background takes the
form
\be
\omega_k^2 \sim \frac{k_3^2}{a^2}+ \left(  \frac{1}{R_{11}}-R_{11} b \right)^2
\equiv \frac{k_3^2}{a^2}+ m_{eff}^2,
\ee
where we have dropped gravitational terms which have been shown to be
of higher adiabatic order and are negligible (recall that these terms scale like $h^2$, and
$h^2 \ll m_{eff}^2$).
Using this frequency we find that the energy density at $t_0$ is
given by
\beq
\rho(t_0)= \frac{\pi^2}{9\vol}  \int_0^\infty \frac{d^3k}{(2\pi)^3}
\sqrt{\frac{k_3^2}{a^2}+ m_{eff}^2} \;
e^{ -\frac{3\pi}{h \gamma} \left( k_3^2+ \frac{1}{2}h^2 \right)}.
\eq
At a later time $t$, the energy density scales in the same way as
the number density,
\beq
\label{rhogo}
\rho(t)=\frac{V(t_0)}{V(t)} \frac{a \, \gamma \,h^{9/2} }
{144 \pi}  \;  m_{eff}^2  \; e^{-\frac{3\pi}{2\gamma}\left( h-a^2 h^{-1}
m_{eff}^2\right)} \mathcal{K}_1\left(  a^2 h^{-1}m_{eff}^2 \right),
\eq
where $\mathcal{K}_1$ is a Bessel function of the second kind.

\subsection{Cosmological Evolution and Trapping}

We are interested in the dynamics after passing through the ESP at $R_{11}=b=1$.
Notice that in the absence of matter the background solution (\ref{FullScaleFactors})
predicts that the radii (moduli) will continue to evolve to larger values.
We will demonstrate that, by including the states produced at the ESP via their stress energy
tensor (\ref{st}),
the motion of the radii will be reversed back towards the ESP, and the moduli can eventually
become trapped.
We again note that there should also be a contribution to the stress tensor coming from the flux,
but such terms scale as $\rho \sim p_i \sim \frac{h^2}{a^6}$ and are therefore only important at
early times (i.e. they are red-shifted by the expansion of $a(t)$).

The evolution begins in the adiabatic region, where $R_{11}>1$, $b>1$ and the mass term
dominates the energy.
In this limit the energy density (\ref{rhogo}) becomes
\be \label{rhoapprox}
\rho \approx  m_{eff} \, n_\chi(t).
\ee
In the opposite limit, near $R,b \approx 1$, the mass is negligible and the energy density becomes
\be \label{rhoapprox2}
\rho \approx \frac{n_\chi(t)}{a}.
\ee
From (\ref{rhoapprox}) we can determine the pressure in the respective dimensions using (\ref{pressures}).
We find that for $R_{11}>1$ and $b>1$ these scale as
\bea
p_{\,3}&\approx&0, \;\;\;\;\;
p_{\,6} \approx-\frac{n_\chi(t)}{6} R_{11}b, \;\;\;\;\;\;
p_{\,11}\approx-n_\chi(t)R_{11}b \;\;\;\;\;\;\;\;
\left( R_{11}>1 \; \text{and} \; b > 1\right).
\eea
We see that for large $R_{11}$ (or large $b$), the negative pressure due to the wrapped M2 branes
dominates.
From the equations of motion (\ref{theeom})
we see that as $R_{11}$ grows large the negative source terms will dominate over
the first terms
on the right side of (\ref{theeom}), given that $H_{11}$ is not too large.
At the turning point $H_{11}=0$ and we find that $\dot{H}_{11}<0$, due to the large negative pressure $p_{11}$,
showing that $R_{11}$ reaches a
maximum value and
turns back toward the ESP \footnote{It is important to note that in anisotropic space-times
negative pressures lead to contraction, not accelerated expansion.}.
After the radii pass through the ESP for a second time, we have $R_{11}<1$, and the
pressures are then given by
\bea \label{pb}
p_{\,3}&\approx&0, \;\;\;\;\;\;\;
p_{\,6}\approx \frac{n_\chi(t)}{6} R_{11}b, \;\;\;\;\;\;\;
p_{\,11}\approx  \frac{n_\chi(t)}{R_{11}} \;\;\;\;\;\;\;\;\;\;\;\;\;\;\;  \left(R_{11}<1\right),
\eea
In the last step we have assumed that the radii have moved sufficiently below the ESP so that we are again
in an adiabatic region, and (\ref{rhoapprox}) still
gives the relevant energy density\footnote{Strictly speaking this is not quite correct.
As the radii pass back through the ESP (non-adiabatic region), further particle production
is possible and the density of particles can increase.  However, including the production of
additional states will only act to enhance the trapping mechanism that we will discuss.
See \cite{Kofman:2004yc} for a related discussion.}.
We see from (\ref{pb}) that as $R_{11}$ continues to evolve towards smaller values, the
pressure becomes large and positive. From
(\ref{theeom}) this means that $R_{11}$ will reach a minimum value and
go back towards the ESP.
This behavior will continue and $R_{11}$ will oscillate around the ESP, with the oscillations
damping due to the expansion of $a(t)$.

However, one place we have been cavalier is in the evolution of the extra dimensional scale factor $b$.
In fact, from its associated
pressure (\ref{pb}) we see that for $b \ll 1$ the pressure will vanish.
Thus, since there is no pressure to
prevent its collapse, $b$ will continue to run to smaller values.
Furthermore, we have over-simplified the entire evolution by assuming that $R_{11}=1$ and $b=1$
occur simultaneously.  If this were the case it would be the result of extreme fine-tuning.
\begin{figure}
\includegraphics[width=10.5cm]{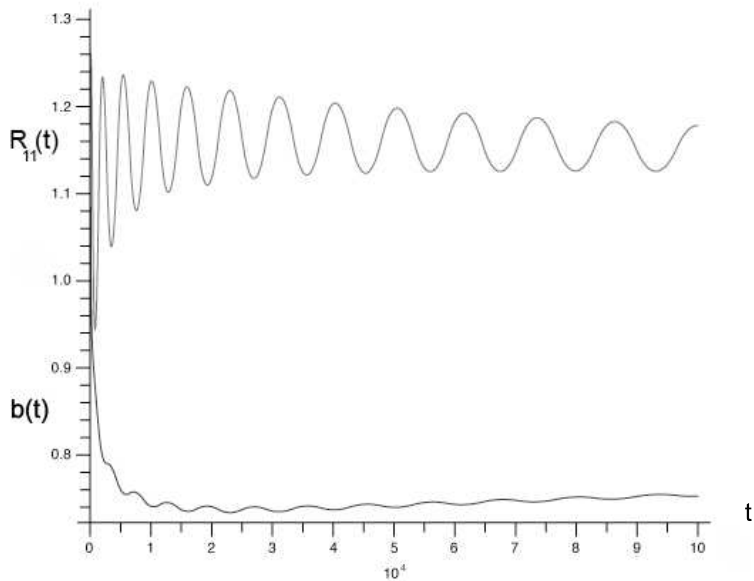}
\caption{\label{fig4} Evolution of $R_{11}$ and $b$ for the exact density (\ref{rhogo}) and pressures, including only the membrane sources related to $R_{11}$.}
\end{figure}
Instead, as can be seen in Fig. \ref{fig4}, we find that for differing values of $R_{11}$ and
$b$ trapping of $R_{11}$ can occur away from the ESP at a value that is determined by the
running of $b$ to its asymptotic value\footnote{We note that upon compactification to
ten dimensions this is the result found in \cite{Watson:2003gf}, where it was shown that the
radii could be stabilized at the expense of a running dilaton}.

Despite the evolution of $b$, we can remedy the situation by simply including the other membrane
states discussed in Section \ref{MthStates}.  That is, near $b=1/\sqrt{R_{11}}$ there are additional
massless states that can be produced with masses
\be
m^{(b)}_{eff}= \left|  \frac{1}{b}-R_{11}(t) b(t) \right|,
\ee
where now the momentum is taken in the $b$ direction (i.e., $k_9 \neq 0$, $k_{11}=0$).
The production of these states is handled analogously
to the previous states, and leads to an additional contribution to the energy density
\be
\rho^{(b)}=6 \, \tilde{n}_\chi(t) \, m_{eff},
\ee
where the factor of six comes from considering states produced equally in all six dimensions.
These states provide the needed pressure term
at small $b$
\bea \label{pressb}
p^{(b)}_6&=& \frac{\tilde{n}_\chi(t)}{b} \;\;\;\;\;\;\;\;\;\;\;\;\;\;\;\;\;\;\;\;
\left( b<1  \right),
\eea
which will cause the motion of $b$ to return to the ESP.

Given the additional sources, we now expect from the asymptotic behavior
that the moduli should be trapped.
However, the dynamics is actually quite involved due to the presence of non-linearities.
Using the experimental result that $H_3>0$ (i.e. we live in three large dimensions),
we examine the system numerically with the results appearing in Figures \ref{fig5} $-$ \ref{fig7}.
In Fig. \ref{fig5}, we see the evolution of the radii  $R_{11}$ and $b$ given initial conditions
consistent with $H_3>0$.
\begin{figure}
\includegraphics[width=10.5cm]{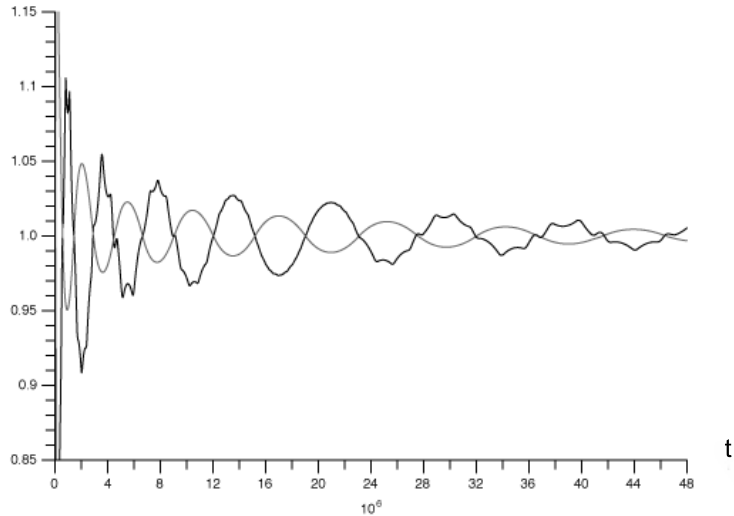}
\caption{\label{fig5} Evolution of $R_{11}$ (dark curve) and $b$ (light curve) for the exact density (\ref{rhogo}) and pressures, including membranes in all extra dimensions.}
\end{figure}
\begin{figure}
\includegraphics[width=10.5cm]{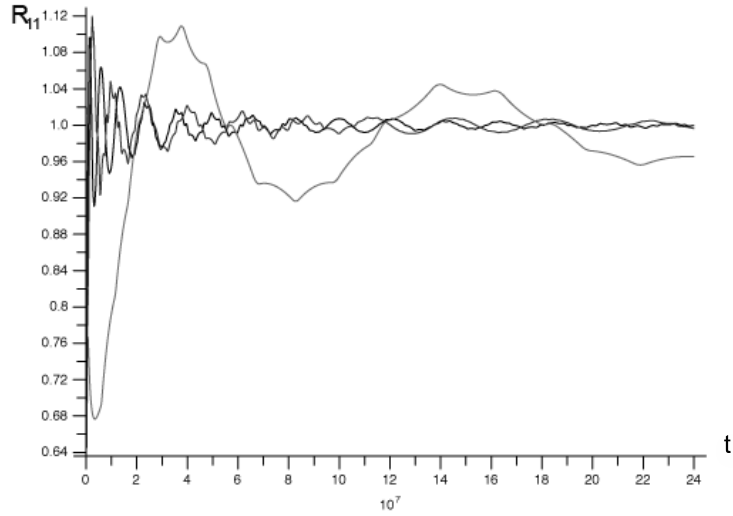}
\caption{\label{fig6} $R_{11}$ for differing initial values of the expansion rate, with $\dot{R}_0=(0.001, 0.5)$ being barely distinguishable and with $\dot{R}_0= 0.6$ (light curve) we see the period begins to grow.  We see that as the initial expansion rate is increased trapping takes longer to occur with $H_{11}>m_p$ leading to the case of no trapping.}
\end{figure}
\begin{figure}
\includegraphics[width=10.5cm]{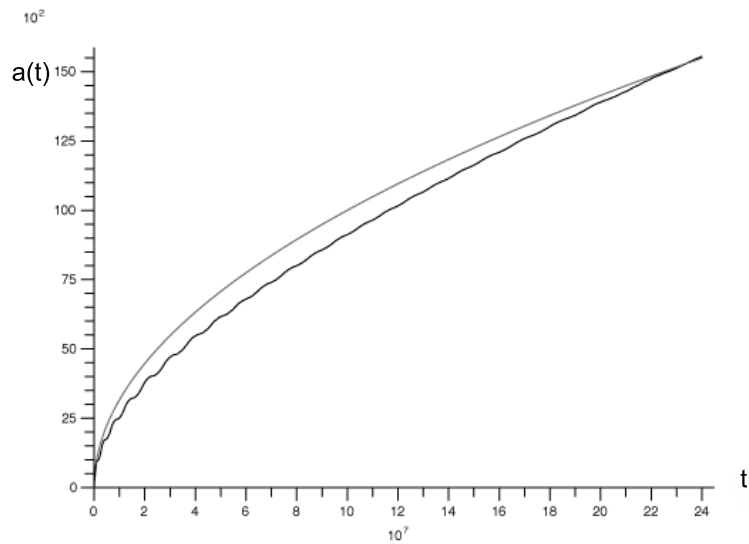}
\caption{\label{fig7} Evolution of $a(t)$ compared to $t^{1/2}$.}
\end{figure}
The jagged oscillations along the curve of $R_{11}$ are not due to numerical error,
but rather
result from the coupling to $b$ and the discontinuities associated with the pressure
changing sign.
We find that the radii will continue to oscillate with a decreasing amplitude, due to
the expansion of $a(t)$.
At late times we find that the radii $R_{11}$ and $b$ approach a constant value, which from (\ref{rhoapprox2}) implies
$\rho \sim a^{-4}$ and $a(t) \sim t^{1/2}$.
That is, our three dimensional universe evolves to that of a radiation dominated universe and
the radii are trapped near the ESP $R_{11}=b=1$.
We find that the trapping is robust, given that the initial expansion rates of the radii do
not exceed the Planck scale (i.e., $H_6<m_p$ and $H_{11}<m_p$.  This condition can be seen
from Fig. \ref{fig6}, where we have plotted the evolution of $R_{11}$ for increasing values of
the initial expansion rate $H_{11}$, and a similar result follows for $b(t)$.
In Fig. \ref{fig7}
we have presented a comparison of the late time behavior of $a(t)$ versus that of a radiation
dominated universe $a(t)\sim t^{1/2}$.  The wiggles in the evolution of $a(t)$ before the radii
completely stabilize, naively may suggest the possibility of cosmological signatures coming from
the trapping mechanism.

\section{Conclusions and Future Prospects}

We have considered both classical and quantum corrections to
low-energy M-theory coming from dynamics of BPS membrane bound
states whose effective mass depends on the radii of the extra
dimensions. Including such states classically leads to an
attractor mechanism that fixes moduli but is inconsistent with the
use of the effective field theory approach.

Insisting on an effective field theory description, we then
consider quantum mechanical production of these membranes in a
time-dependent background, and find the expected result that the
production suffers Planckian suppression. However, we do find the
possibility of significant and finite production for states that
exhibit enhanced symmetry. We believe that these should correspond
to non-threshold bound states of membranes and gravitational
waves, which become tensionless at the eleven-dimensional
self-dual point, $R_{11}=l_p$. An exact construction of such
states is challenging, due to the problems of quantizing the
supermembrane and the lack of an effective description of the
theory in this regime.

In \cite{Russo:1996if} it was shown that one can obtain the BPS
spectrum of the supermembrane in special limits and by utilizing
properties of BPS configurations. However, the enhanced states we
are interested in require non-vanishing vacuum energy, which is
not compatible with the $\mathcal{N}=2$ SUSY case considered in
\cite{Russo:1996if}. Therefore, guided by the analysis of
\cite{Russo:1996if}, we conjecture the existence of the enhanced
states, awaiting a more concrete construction. One possibility for
their existence is the case of $\mathcal{N}=1$ heterotic M-theory,
in analogy with the enhanced gauge symmetry of heterotic strings.
Another intriguing possibility is the recent conjecture that the
spectrum of string / M-theory contains states that lie below the
BPS bound \cite{Arkani-Hamed:2006dz} .

Assuming that such enhanced states exist, we find that they can
have a critical impact on the evolution of moduli, which would
have been missed in the naive low-energy effective theory
neglecting dynamics. We have found that, by including the
backreaction of the membrane states produced, the radii are
dynamically attracted to values near fixed points of S- and T-
duality. This effect would be missed if one did not have a
knowledge of (enhanced) M-theory states having masses which depend
on evolving moduli. Thus, the lesson we learn is that the presence
of time-dependence introduces a dynamical mass scale that must be
taken into careful consideration in the effective field theory.
Furthermore, it is paramount not to forget the string theory
origin of the low-energy effective action.

In addition to the corrections from on-shell particle production
which we have explored here, there will be radiative corrections
coming from the presence of the light membrane states. In fact, it
was shown in \cite{Silverstein:2003hf} that, for the case of
colliding D-branes, open strings becoming light lead to
corrections to the gauge theory propagator resulting in a speed
limit for the moduli. In the case we have considered here a
similar story should hold, but this time it is the closed string
moduli which should be slowing down (the radii of the extra
dimensions). This offers a challenge, since the gauge theory
interpretation of such processes is unclear. We find the
possibility of speed limits for radii intriguing, and we hope to
report on it shortly.

\begin{acknowledgments}
We would like to thank R. Brandenberger, B. Holdom, L. Kofman, A. Krause, D. Lowe, L. McAllister,
S. Patil, D. Podolsky, S. Prokushkin, and H. Nastase for useful discussions.  We would especially like to
thank A. Jevicki, J. Russo, and A. Tseytlin for critical comments and suggestions and we are
grateful to K. Hori for finding an error in an earlier draft.  S.C. would like to thank the
University of Toronto for hospitality.  This work was financially supported in part by the
National Science and Engineering Research Council of Canada, the Schnabel Woods institute,
and the U.S. Department of Energy under contract DE-FG02-91ER40688, TASK A.
\end{acknowledgments}

\end{document}